\documentclass[preprintnumbers,amsmath,amssymb,superscriptaddress,twocolumn,showpacs]{revtex4-1}
\usepackage{graphicx}% Include figure files
\usepackage{dcolumn}% Align table columns on decimal point
\usepackage{bm}% bold math
\usepackage{natbib}
\usepackage{physics}
\usepackage[caption=false]{subfig}
\usepackage{color}

\newcommand{\be}{\begin{equation}}
\newcommand{\ee}{\end{equation}}
\newcommand{\bea}{\begin{eqnarray}}
\newcommand{\eea}{\end{eqnarray}}

\newcommand*{\hham}{\hat{\mathcal{H}}}

\begin{document}

\title{Angle locking of a Levitating Diamond using Spin-Diamagnetism}

%Magnetic-optical tuning of the diamond cristalline axis 

\author{M. Perdriat$^1$, P. Huillery$^1$, C. Pellet-Mary$^1$, G. H\'etet} 
\affiliation{Laboratoire De Physique de l'\'Ecole Normale Sup\'erieure, \'Ecole Normale Sup\'erieure, PSL Research University, CNRS, Sorbonne Universit\'e, Universit\'e de Paris , 24 rue Lhomond, 75231 Paris Cedex 05, France}

\begin{abstract}
Nano-diamonds with embedded nitrogen-vacancy (NV) centers have emerged as promising magnetic field sensors, as hyper-polarizing agents in biological environments, as well as efficient tools for spin-mechanics with levitating particles. These applications currently suffer from random environmental interactions with the diamond, such as gas collisions, which implies poor control of the N-V direction and reduces its polarization drastically. Here, we predict and report on a strong diamagnetism of a pure spin origin using a population inversion close to a level crossing in the NV center electronic ground state. Using this effect, we show angle locking of the crystalline axis of a trapped micro-diamond along an external magnetic field with bright perspectives for these applications. \end{abstract}

\maketitle

The electronic spin of the negatively charged nitrogen-vacancy (NV) in diamond has been employed in a broad range of research directions over the past two decades with impressive demonstrations in nanoscale magnetometry \cite{Casola}, and in quantum communication and computing \cite{Awschalomrev}. 
%The results in these areas were made possible thanks to the unique optical read-out and polarization capabilities offered by the NV center \cite{hensen, degen review}. 
In the employed platforms, high quality diamonds are typically tethered to a scanning tip for imaging fields above magnetic materials or to a cold finger for minimizing phonon-induced relaxation in coherent spin manipulations. %They also generally require CVD-grown bulk diamonds for optimized spin coherence properties.
%Using fixed diamonds offers the possibility to optimize their properties for the required tasks as well as to readily align magnetic field to the NV axes. 
Conversely, major other applications involving the electronic spin of the NV center require diamonds to be untethered. 
In particular, freely moving or loosely bound nanodiamonds containing NV centers can serve as {\it  in vivo} hyper-polarizing agents in nuclear magnetic resonance \cite{McGuinness, Miller, Waddington, QuanLi} or as nano-magnetometers in living cells \cite{Schirhagl}.
%Most experiments have been realized with bulk diamonds \cite{BucherPRXWals, Broadway, Acebal}, 
%but 
%optically detected magnetic resonance {\it in vitro} \cite{McGuinness, Miller}, polarization of nuclear spins in solution \cite{Waddington} as well as the 6D tracking of the motion of a cell \cite{QuanLi} with individual nanodiamonds offer a clear path towards {\it in vivo} NMR with NVs.
%For optimum performance as well as for practical use, the technique should be performed with nanodiamonds immersed in liquid \cite{McGuinness}.
%Using bulk tethered diamonds does not lend itself to {\it in vivo} NMR using NV centers.. 
Another example is spin-mechanics with levitating particles \cite{PerdriatRev}. There, NV centers are employed to create quantum superpositions of the diamond motion. The underlying idea is that quantum superpositions of the NV electronic spin states can be transferred to the diamond center of mass motion using magnetic field gradients \cite{rabl}, which offers perspectives for matter-wave interferometry and tests of quantum gravity \cite{Pedernales, yin, scala, wan} . 
%NV centers in trapped diamond mechanical oscillators could also be deployed for observing quantum phase transitions  \cite{LMG tongcang, BO ? } 
%When levitating, nanodiamonds are subjected to air flows or laser fields that make them move over the course of the measurements. 

%One major difficulty for most applications involving NV centers is to align the magnetic field to the NV axis. Off axis magnetic fields indeed lower the polarization when magnetic field are above 100 G.  
When operated in a liquid or trapped under vacuum however, the orientation of the nanodiamonds are subjected to various conservative or viscous forces \cite{DelordPRL, Geiselmann, Neukirch2, Horowitz}. The spin direction thus changes over time, making it difficult to make full use of its capabilities for sensing and coherent control. One major concern is the lack of NV polarization when the external magnetic field is not along the diamond $\langle 111\rangle$ axis \cite{tetienne4}.
%It would thus be highly beneficial to be able to control the diamond angle at a distance.
%It was shown that the diamond orientation held in position by electrostatic torques, could be deviated slightly from its equilibrium position when a microwave tone is scanned close to the NV center's spin resonances. 
%Measurements of the magnetization of NV-doped diamonds is still lacking. 
%To enter these regimes, the magnetic response of the NV center must be fully controlled and understood.  
Here, we show that NV-doped diamonds behave as materials with strong magnetic anisotropy and can be turned into diamagnets with a large susceptibility.
Further, we exploit the spin-diamagnetism to stabilize the $\langle 111\rangle$ crystalline axes of levitating micro-diamonds along the applied magnetic field. 

To understand the origin the NV-doped diamond magnetism, let us first consider a generic non-magnetic crystal hosting independent atomic defects with unpaired electrons. The dominant magnetization of such a crystal is quantified through the variation in the electronic energy levels $\epsilon_k$ of the defects as a function of an applied magnetic field $\bm B$ via the formula:
\begin{eqnarray}
M_i=-d \sum_{k}\frac{\partial \epsilon_k}{\partial B_i}p_k,
\end{eqnarray}
where $d$ is the density of the defects, $B_i$ is the magnetic field component along the $i$ direction and $p_k$ is the population in the state $k$.
Note that we neglect the demagnetizing field from the defects, which is much smaller than the applied field in typical dilute spin ensembles. The components $\chi_{ij} = {\mu_0}\frac{\partial M_i}{\partial  B_j}$ of the magnetic susceptibility tensor $\underline{\underline{\chi}}$ then read:\begin{eqnarray}
 \chi_{ij}=-d \mu_0 \big[ \sum_{k}\frac{\partial \epsilon_k}{\partial B_i} \frac{\partial p_k}{\partial B_j} + \sum_{k}\frac{\partial^2 \epsilon_k}{\partial B_i \partial B_j}p_k \big]. 
\end{eqnarray}
The first term in this equation is positive, strongly temperature dependent and gives rise to Langevin paramagnetism.
%where $\underline{\underline{\chi}}$ follows the Curie's law \cite{?}.
The second term typically has two opposing contributions: a negative contribution coming from the orbital electron motion, namely the Larmor diamagnetism, and a positive Van Vleck paramagnetic contribution \cite{van1932theory, carlin2012magnetochemistry}. As we will show, close to a level crossing, NV-doped diamonds in fact exhibit strong Van Vleck magnetism with a tunable susceptibility.
   \begin{figure*}[htbp]
\centering
 \includegraphics[width=\linewidth]{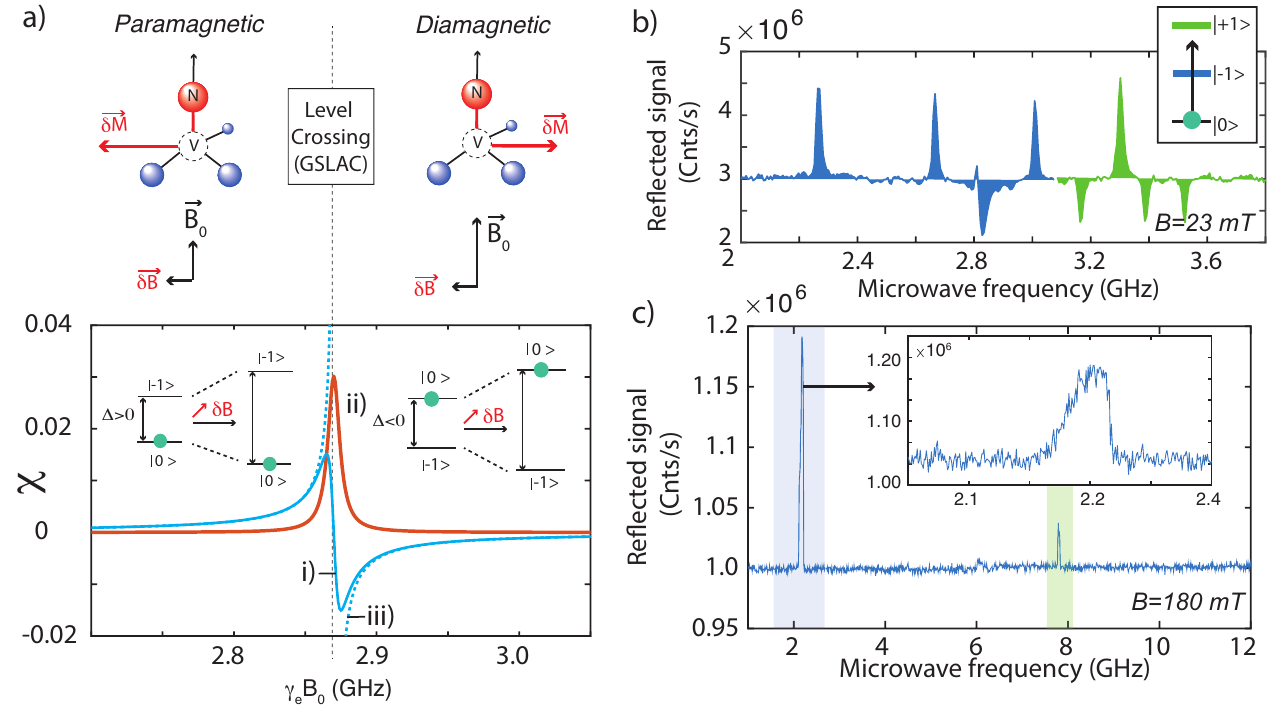}
\caption{a) Trace i), resp. trace ii), show the expected NV Susceptibility coefficients $\chi_\perp$ resp. $\chi_d$ as a function of $\gamma_e B_0$. Trace iii) (in dashed lines) displays $\chi_\perp$ obtained from second order perturbation theory. b) and c) are Mechanically-Detected-Magnetic-Resonances measured in the paramagnetic (resp. diamagnetic) regime where B=23 mT and 180 mT respectively. The inset of c) is a zoom on the first spin-resonance at 2.2 GHz.}
\label{fig1}
\end{figure*}

The ground electronic state of the NV$^-$ center is a triplet state with a zero-field splitting $D$ coming from magnetic dipole-dipole interaction between two unpaired electrons \cite{Doherty}. This interaction sets a natural quantization axis $z$ along the N-V direction shown in the Fig.\ref{fig1}-a). 
%We will show that the magnetic response of the NV$^-$ center obeys Van Vleck magnetism, exhibiting both paramagnetism and diamagnetism. %
%
One feature of the electronic spin of the NV$^-$ center is that it can be optically polarized to the $\vert m_s=0 \rangle$ state using green laser light.
%a green laser light can polarize the electronic spin in the $\vert m_s=0 \rangle$ state \cite{Doherty}. 
In the absence of magnetic field, the other $\vert m_s=\pm 1 \rangle$ states both lie $D\approx(2\pi) 2.87$ GHz above the $\vert m_s=0 \rangle$ state at room temperature. %
In the presence of a magnetic field $\bm B_0$ aligned with the NV axis, the Hamiltonian of the NV$^-$ center electronic spin reads:
\begin{eqnarray}
\hham_0=\hbar D\hat{S}_z^2+\hbar \gamma_e  B_0 \hat{S}_z, \end{eqnarray} 
 where $\gamma_e$ is the electron gyromagnetic ratio.
In this magnetic field configuration, there is no induced magnetization of the NV-doped diamond.

As is the case with several paramagnetic crystals \cite{abragam, van1932theory, White}, the essence of the NV magnetization lies in the state mixing induced by a static magnetic field perturbation $\delta B_\perp$ that is not along the quantization axes defined by the crystal field.  Such state mixing generates a magnetic moment that is perpendicular to the NV axis because of the mixing between the $\vert m_s=0\rangle$ and $\vert m_s=\pm 1\rangle$ states. Adding a static perturbation $\bm {\delta B}$, the Hamiltonian becomes $\hham=\hham_0+\hat{V}$ where $\hat{V}=\hbar \gamma_e  \bm {\delta B} \cdot \bm {\hat{S}}$ so that the NV$^-$ center acquires a magnetization $\bm {\delta M}= \underline{\underline{\chi}}^{NV}\bm {\delta B}/\mu_0$. 
%We propose to calculate $\underline{\underline{\chi}}^{NV}$
 $\underline{\underline{\chi}}^{NV}$ is rotationally invariant along the $\bm{e_z}$ axis so
%In the orthogonal basis $(\bm{e_x},\bm{e_y},\bm{e_z})$ where $\bm{e_z}$ is the NV direction, $\underline{\underline{\chi}}^{NV}$ is rotationally invariant along the $\bm{e_z}$ axis. 
its only non-zero components are $\chi_\perp=\chi_{x,x}=\chi_{y,y}$ and $\chi_d=\chi_{y,x}=-\chi_{x,y}$ (see Sec. I of the Supplementary Material (SM)\cite{SM_spin_diam}\nocite{HuilleryCPT}).

%We want to calculate these two coefficients. 
 To quantify $ \underline{\underline{\chi}}^{NV}$, we first consider a single NV center orientation out of the four $[111]$ directions, take a density $d = N/V \approx 1$ppm and suppose an optical pumping rate $\gamma_{\rm las}$ to the $\ket{m_s=0}$ state that is far greater than the longitudinal spin relaxation rate $T_1\approx 500 \mu$s \cite{Pellet}. 
%The most dominant susceptibility term is our experiment is $\chi_{\perp}$, which reads
%
%\begin{eqnarray}\label{Mx2}
%\chi_\perp = d  \hbar \mu_0  \sum_{i=-1,1} \frac{\gamma_e^2\Delta_{i}}{\Delta_{i}^2+(\Gamma_2^*)^2}, \end{eqnarray}
% where $\Delta_{i}=D+i\gamma_e \mu_0 H_0$ and $\Gamma_2^*/2\pi = 1/T_2^*\approx 5$ MHz is the spin dephasing rate due to the fluctuating dipolar coupling to substitutional nitrogen atoms (see Sec. I of the SM \cite{SM_spin_diam}). 
$\chi_\perp$ resp. $\chi_d$ are plotted in Fig. \ref{fig1}-a). 
%displaying a dispersive resp. a lorentzian profile.
 %For the rest of the demonstration, we will consider that $\Gamma_2^* \ll \Delta_i$ so that $\chi_d$ can largely be negligible compared to $\chi_\perp$.
% 
% 
%\begin{eqnarray}
%\chi_d = d  \hbar \mu_0  \sum_{i=-1,1} \frac{\gamma_e^2i \Gamma_2^*}{\Delta_{i}^2+(\Gamma_2^*)^2}(p_0-p_i). \end{eqnarray}
%
%For a sufficiently strong optical pumping process in the $\ket{m_s=0}$ state, we can consider that $p_0 \approx 1$ and $p_1=p_{-1} \approx 0$. 
In the limit where $\gamma_e  B_0 \ll D$, $\chi_{\perp} \approx d \hbar \mu_0  2\gamma_e^2/D\approx 10^{-4} >0$ so that the NV-doped diamond is strongly paramagnetic.
Including the four NV centers orientations along the four $[111]$ diamond crystalline axes now, the total susceptibility drops to about $10^{-5}$. This is comparable to the diamond orbital diamagnetism ($\chi_{\rm orb}\approx  10^{-5}$) \cite{Pellet}, which is negligible in our experimental conditions. Close to the ground state level anti-crossing (GSLAC) however, where $\gamma_e B \approx D$, the states $\vert m_s=0\rangle$ and $\vert m_s=-1\rangle$ mix much more. In this regime, one for instance finds: \begin{eqnarray}
\label{Mx}
 \chi_\perp &\approx&d \hbar \mu_0  \frac{\gamma_e^2 \Delta}{\Delta^2+(\Gamma_2^*)^2},
 \end{eqnarray}
 where $\Delta=D-\gamma_e  B_0$ and $\Gamma_2^*/2\pi = 1/T_2^*\approx 5$ MHz is the spin dephasing rate due to the fluctuating dipolar coupling to substitutional nitrogen atoms (see Sec. I of the SM \cite{SM_spin_diam}). 
The formula for $\chi_\perp$ and $\chi_d$ agree very well with the numerical calculations used in Fig. 1a).
Importantly, the maximum susceptibility $\vert \chi_\perp \vert \approx 10^{-2}$ is two orders of magnitude larger close to the GSLAC than far off-resonance. Further, when $\Delta <0$ and in the dispersive limit $\vert \Delta\vert > \Gamma_2^*$, the NV spin population is in the first excited state (see inset of Fig. 1a)). The NV intersystem crossing in the optically excited state $^3 E$ indeed preserves polarization in the $\vert m_s \approx  0\rangle$ state even after the GSLAC, which results in a population inversion, as in the recently demonstrated diamond maser \cite{Breeze}. 
The possibility to swap populations between $\vert m_s \approx  0\rangle$ and $\vert m_s \approx  -1\rangle$ states using $\Delta$ thus results in a tunable magnetization that can be opposed to (as in a diamagnet) or aligned with (as in a paramagnet) the applied magnetic field perturbation. 

We can interpret the NV$^-$ induced magnetization in terms of Van Vleck magnetism using perturbation theory away from the level crossing. The  modification of the energies $\epsilon_i$ of the three NV$^-$ eigenstates $\ket{i}$ due to $V$ are given by:
\begin{eqnarray}
\epsilon_i - \epsilon_i^{(0)} \approx \bra{i}\hat{V}\ket{i}+\sum_{j \ne i} \frac{|\bra{i}\hat{V}\ket{j}|^2}{\epsilon_i^{(0)}-\epsilon_j^{(0)}}
\end{eqnarray}
to leading order.
The first Langevin term is negligible in the case of magnetic field perturbation that is perpendicular to the NV axis.
Since the orbital momentum of the NV$^-$ center is zero in the ground state, there is also no Larmor diamagnetism. Just like in Van Vleck magnetism, the dominant contribution to the susceptibility therefore comes from the second order term which, using Eq. $2$, yields:
\begin{eqnarray}
 \chi_{V.V.} = d  \hbar \mu_0  \sum_{i=-1,1} \frac{\gamma_e^2}{\Delta_{i}}(p_0-p_i).
 \end{eqnarray}
This expression perfectly matches the Eq. \ref{Mx} derived close to the GSLAC, if we set $\Gamma_2^* \rightarrow 0$, $p_0\gg p_{-1}$ and $\Delta=\Delta_{-1}\ll \Delta_{+1}$. Note that in all reported magnetic materials thus far, the spin population resides mostly in the lowest energy state, following Boltzmann's distribution.  Instead, in the NV case, an excited state of the system can be populated in the steady state, which leads to a diamagnetic behavior where $\chi_\perp <0$.

 \begin{figure}[htbp]
\centering
\includegraphics[width=\linewidth]{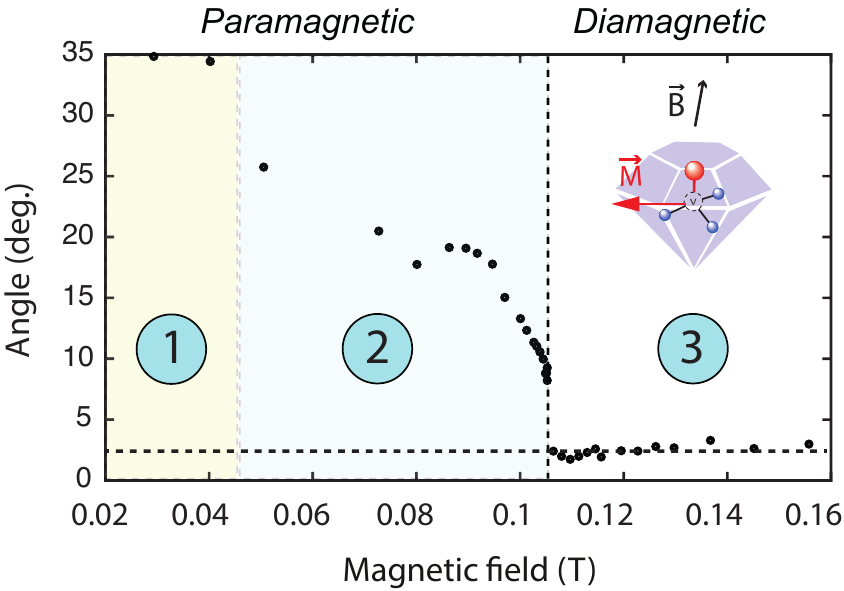}
\caption{Angle of one of the diamond $[111]$ axes versus magnetic field amplitude.}
\label{fig2}
\end{figure}

One way to determine the magnetization $\bm M$ of a material is to measure the magnetic torque $\bm \tau= V \bm M(\bm B) \cross \bm B$ applied to it under an external magnetic field $\bm B$. For NV$^-$ centers at a small angle $\theta$ with respect to $\bm B$, we can write $\bm {B} \approx B_0 \bm{e_z}+\theta B_0 \bm{e_x}$. Treating the transverse field $\bm {\delta B}=\theta B_0 \bm{e_x}$ as a small magnetic perturbation, we get $\bm \tau= (V/\mu_0) \chi_\perp B_0^2 \theta \bm{e_y}$.
As expected, if $\chi_\perp$ is negative (resp. positive), the torque tends to rotate the particle so that the NV center axis is aligned (resp. perpendicular) to the magnetic field. 

In our experiment, a micro-diamond with a concentration of NV centers between 3-5 ppm is loaded in an electrostatic trap with kHz angular frequencies (see Sec. II of the SM \cite{SM_spin_diam}). 
The combined action of the Paul trap and magnetic torques $\bm \tau$ determines the angle of the four NV centers with respect to the external magnetic field.
In order to characterize $\bm \tau$ and thus the NV-induced magnetism, we apply green laser light and sweep the frequency of a microwave signal at a power that is far from saturating the spin-transition. At a spin resonance, a small magnetization is generated along $\bm {e_z}$, at which point the equilibrium angular position of the diamond is slightly modified. This MDMR technique \cite{DelordNat} thus enables tracking of the angles of the four NV orientations with respect to the magnetic field $\bm B$. 
In order to measure $\bm \tau(\bm B)$ and to probe the spin-diamagnetism, we then run several scans under different magnetic field strengths (see Sec. III and VI of the SM \cite{SM_spin_diam}). 

 \begin{figure}[htbp]
\centering
\includegraphics[width=\linewidth]{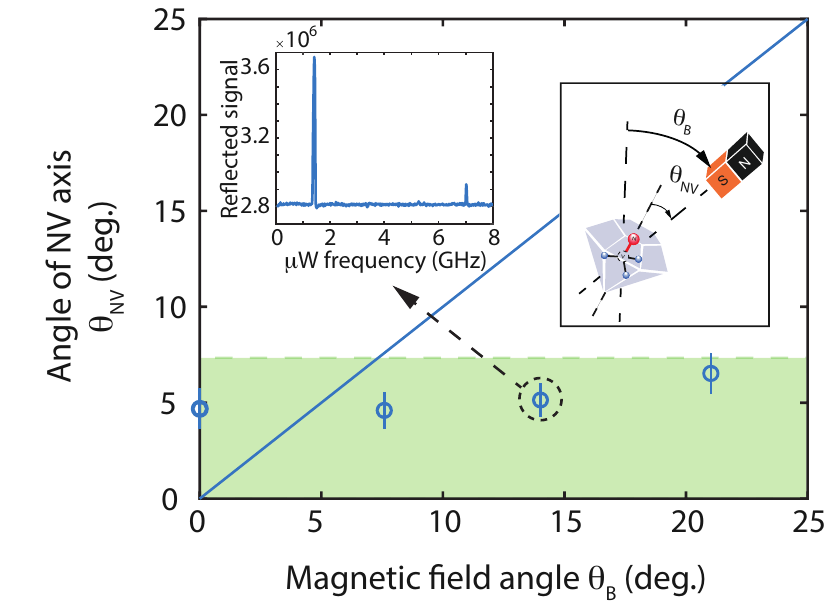}
\caption{Angle of the diamond $[111]$ axis $\theta$ versus magnetic field angle $\theta_B$ (blue circles). See sketch in the inset on the top-right. The plain blue line is the response that would be obtained without NV induced magnetism. The inset on the top-left is a mechanically-detected magnetic-resonance scan taken at $\theta_B\approx 14$ degrees.}
\label{fig3}
\end{figure}

One of the spectra is plotted in Fig. \ref{fig1}-b) for $B\approx 23$mT, showing the characteristic opposite signals for $\ket{0}$ to $\ket{\pm 1}$ transitions of the same NV orientation  \cite{DelordNat}. Several other MDMR scans have been realized with $B$ fields varying from 10 to 40 mT. It was found that the extracted angles of the four NV orientations do not depend significantly on the applied magnetic field (see Sec. IV of the SM). The equilibrium angular position is thus largely determined by the electrostatic trap angular potential. 
A spectrum taken at $B\approx 180$~mT is shown in Fig. \ref{fig1}-c). Here, only two lines clearly detach from the spectrum with a very large signal to noise ratio. Importantly, their frequencies are consistent with an almost ideal alignment of one NV axis along the magnetic field.  
Such radically different spectra when $B\gtrapprox105$~mT, are observed in almost all of the particles that we trap, pointing towards their microwave-free magneto-optical rotations. 
A closer look to the $\vert m_s=0 \rangle$ to $\vert m_s=-1 \rangle$ transition also confirms this. Indeed, the blue-detuned sharp edge of the mechanical response, observed in the inset of Fig. \ref{fig1}-c), is opposite to what is observed in the $\gamma _e B \ll D$ regime \cite{DelordNat} (on the red side),  indicating that reducing population in the $\vert m_s\approx 0 \rangle$ state using the microwave misaligns the NV and magnetic field axes (see Sec. V of the SM \cite{SM_spin_diam}). 
 \begin{figure}[htbp]
\centering
\includegraphics[width=\linewidth]{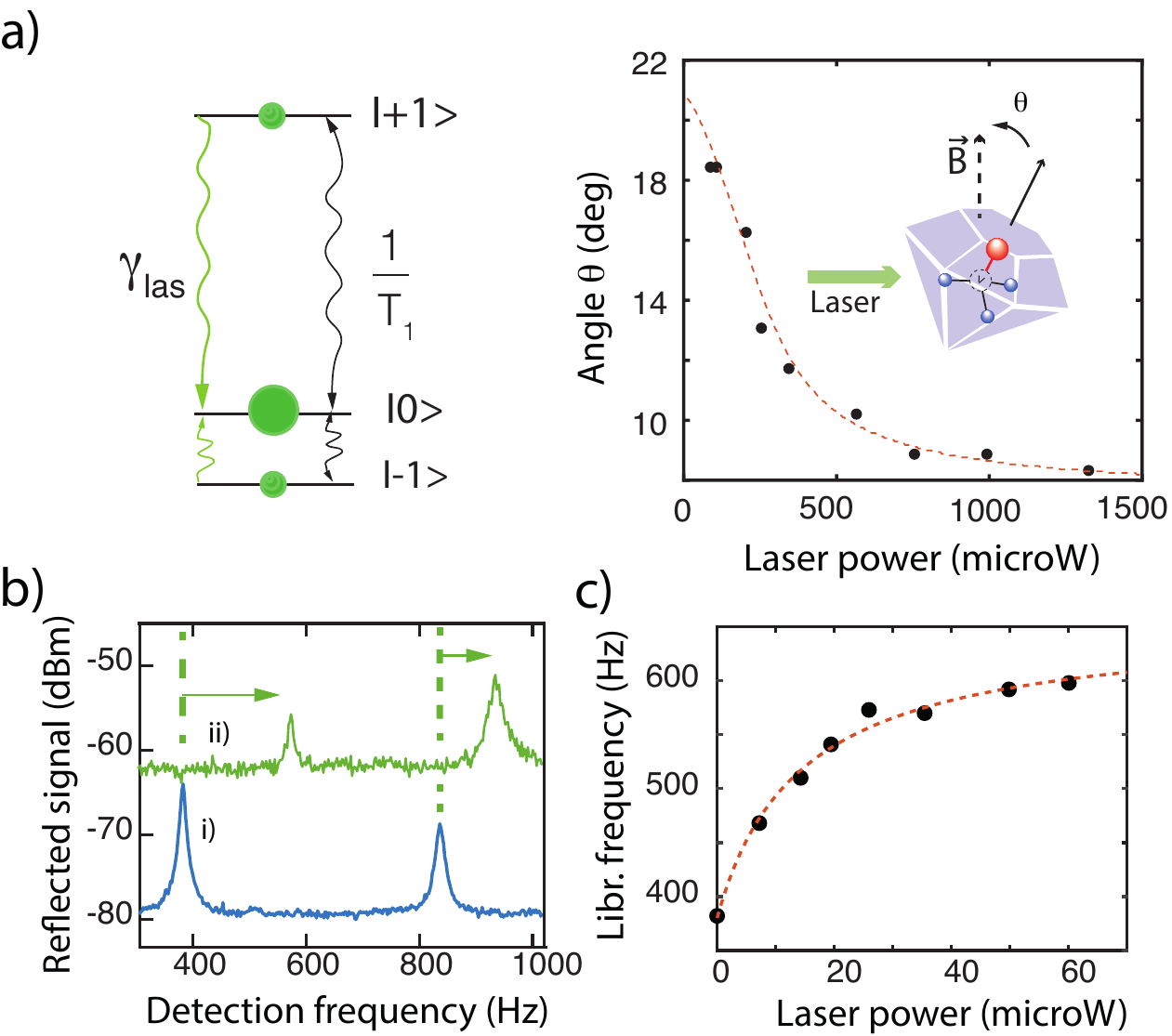}
\caption{a) Left: NV center's electronic spin eigenstates in the $^3A_2$ state. Green arrows depicts the effective optical pumping to the $\vert m_s=0\rangle$ state that results from the intersystem crossing in the optically excited $^3 E$ state (not shown). Black arrows depict depolarizing mechanisms at rates $1/T_1$.  Right: angle of one of the NV axes with respect to the magnetic field direction (as shown in the inset) as a function of the green laser power. b) Trace i), resp. trace ii), show the power spectral density of the librational modes without/with the green laser respectively. c) Librational frequency of the first mode as a function of laser power.}
\label{fig4}
\end{figure}

Numerical calculations of the magnetic energy of the NV-doped diamond are shown in the section VII of the SM \cite{SM_spin_diam}).
The four $[111]$ directions of the NV centers are calculated to be strongly confining, with a potential depth that is three orders of magnitude larger than $kT\approx 10^{-21}$J at T=300 K, on the order of the Paul trap confinement. Beyond the GSLAC, the spins can thus efficiently pull the [111] crystalline direction towards the magnetic field direction, as expected for a diamagnetic anisotropic material.

In Fig. \ref{fig2}, the angle $\theta$ of the NV centers orientation that is closer to the magnetic field axis is plotted as a function of the magnetic field amplitude (see Sec. IV of the SM \cite{SM_spin_diam} for details about the angle calibration technique). Three regions can be identified. When the magnetic field ranges from 0 to 45 mT (region 1), $\theta$ does not change significantly.  When the magnetic field increases from 45 mT to 105 mT however, $\theta$ reduces until it suddenly drops. The dependence of $\theta(B)$ in regions (1) and (2) results from a competition between the torque applied by the Paul trap and by the other three NV centers orientations (see Sec. IV of the SM).
Importantly, close to 105~mT, a transition occurs where the angle between the NV centers and the magnetic field reaches $\approx 2^\circ$. The NV orientation remains locked to this value over a wide range of magnetic fields.
The overall change of the NV angle is thus consistent with the above described transition from a paramagnetic state to a diamagnetic state at the ground state level anti-crossing.
The latter transition is expected to take place at $B_c=$102.4~mT, which is very close to the observed critical B-field ($B_c \approx$ 105 mT). We explain this difference by the residual Paul trap torques which shift both $\theta$ and $B_c$ (see Sec. VI of the SM).
Another check that firmly establishes NV spin induced diamagnetism is that the particle [111] direction follows the magnetic field direction. This is demonstrated in Fig. \ref{fig3} where, as the magnetic field direction is rotated, the angle between the NV and magnetic field axis departs from the linear law and remains at small angle values. Note that these are typical observations. Some diamonds with larger NV concentrations were even found to reach 
$\theta$ values that were less than a degree.

One of the striking feature of the observed diamagnetism is its optical tunability. 
%
%The magnetic response is particularly enhanced by the laser induced spin-polarization (see Fig. \ref{fig4}-a)) which reduces the effective spin-temperature close to the $\mu$K. 
Fig. \ref{fig4}-a) shows that the NV axis can indeed be steered towards the magnetic field thanks to the laser induced polarization to the $\ket{m_s=0}$ state.
%The observed optically controlled magnetization of the doped diamond overwhelms
%the diamagnetic contribution from the orbital motion of the electrons in the diamond. 
The magnetically-induced librational frequency is also expected to increase with laser power once in the diamagnetic regime (region (3)). 
%Using Eq. \ref{Mx} and the expression for  $\bm \tau$, one finds that it indeed overwhelms the Paul trap confinement in the strong optical power regimes.
Fig. \ref{fig4}-b) i and ii) show the power spectral density (PSD) of the librational modes without (i) and with (ii) the green laser, showing a clear frequency shift of both librational modes (see Sec. VI of the SM \cite{SM_spin_diam}). The increase of the observed librational frequencies with laser power is plotted in Fig. \ref{fig4}-c), showing again the strong magneto-optical nature of the effect. A study of its dependence with the magnetic field also shows the resonant boost of the librational frequency close to the GSLAC (see Sec. VII of the SM).

Our observations open up several intriguing directions. 
First, the discovered diamagnetism should be strong enough to observe the diamagnetic {\it force} from the NV centers.
% similar to what was achieved using the inherent diamond diamagnetism due to the electron motion \cite{Hsu, OBrien}. 
Calculations even suggest that increasing the NV spin density by a factor of 10 and working with purer samples \cite{Edmonds, TALLAIRE2020421, MINDARAVA2020182} will create an internal magnetization that is of the order of the applied magnetic field offering the prospect of using a diamond to expel magnetic field lines and thus to levitate them like in superconductor levitation.

%Note that Spin-diamagnetism may also be observed in systems that can be optically inverted, such as the $VH-$ in hBN or SiC. (CITE)
Being able to align diamond main crystalline axes along magnetic fields also means that the embedded NVs can be easily employed for polarizing other species, such as the nuclear spins of the $^{13}$C atoms of molecules under liquid environments \cite{Shagieva, Broadway, Acebal}. Many groups are currently working towards polarizing paramagnetic species in solution using the NV centers inside nanodiamonds \cite{Waddington}. One difficulty with current attempts is to achieve large NV polarization under a fluctuating environment and under necessarily strong magnetic fields that are well-aligned to the NV axis. 
%Notably, cross polarization is efficient close to the GSLAC \cite{WangBajaj}, when the B-field is tuned close the NV axis. 
Our technique thus counteracts this drawback and may thus boost the hyper-polarization efficiency \cite{Tetienne3}, opening a path towards more efficient NMR.

More generally, our work may find applications for local remote control of the motion of nano- or microscopic objects. 
%Typically, the remote control of the motion of micro-crystals is done quite effectively using lasers \cite{Geiselmann}, magnetics \cite{Abbott2, Chuang, Liu} or electric fields, as in Microelectromechanical systems (MEMS). 
%Local remote control of the orientation of an object at the nanoscale is
%often a very difficult task. %Thus far only micron-sized objects \cite{delordPRL} were shown to enable $T_2^*$-limited ODMR lines in a magnetic field. 
As an example, the present magneto-optical confinement offers a solution for micro-wave-free stabilization of the angle of NDs in optical tweezers \cite{Geiselmann, Horowitz, Neukirch2} with important perspectives for further spin-mechanical experiments in the quantum regime \cite{Gonzalez, Millen, Ma, BoboWei, wan}. \\

{\bf Acknowledgments}
We would like to thank L. Rondin, V. Jacques, A. Dréau and C. Voisin for fruitful discussions, as well as Ludovic Mayer for lending us microwave equipment. 
GH acknowledges SIRTEQ for funding.

\newpage

$~$\\

\begin{widetext}

\vspace{0.2in}
{\Large \hspace{2.13in}\textsc{Supplementary Material} }\\
\

In this supplementary material, we provide an analytical treatment of the induced magnetization and of the torque exerted by the spin of the NV centers in the trapped diamond. We also present extra experimental details on angle calibrations as well as explanations about the angular response in the non-linear regime. 

\tableofcontents

\section{Calculation of the NV$^-$ magnetic susceptibility tensor}

In this I, we will consider the simplified regime where NV centers have a quantization axes that make a small angle with respect to the magnetic field. As we will show, after the GSLAC, the doped-diamond is in the diamagnetic phase, where one of the [111] crystalline diamond directions is aligned to the B field.
In this limit, the magnetic field response for that class is an order of magnitude larger than that from the other classes of NV centers who are largely mixed by the transverse magnetic field so this approximation is well-justified. 

Because of the many uncertainties in the experiment, such as on the number of spins, on the moment of inertia and main confinement axis due to the quadrupolar charge distributions, the theory is meant to provide a qualitatively description of the effects.

We calculate the magnetic response $\bm {\delta M}$ of a NV$^-$ center under a static longitudinal magnetic field $\bm {B_0} = B_0 \bm {e_z}$ and a small magnetic perturbation $\bm {\delta B}$. The magnetization $\bm M=- d \hbar \gamma_e \langle \hat{\bm S} \rangle$, where $\gamma_e=g_e e/(2m)$ is the electron gyromagnetic ratio, taken as being positive here. The magnetic susceptibility tensor is defined as the relationship between the magnetic response $\bm {\delta M}$ of a material and a magnetic field perturbation $\bm {\delta B}$ such that $\bm {\delta M}=\underline{\underline{\chi}}^{NV} \bm {\delta H}$. The NV Hamiltonian equals $\hham=\hham_0+\hat{V}$ where:
\begin{align}
\hham_0 & =\hbar D\hat{S}_z^2+\hbar \gamma_e  B_0 \hat{S}_z \quad  {\rm and} \quad \hat{V} = \hbar \gamma_e  \bm {B} \cdot \hat{\bm S}.
\end{align}
The first term accounts for the anisotropy of the induced magnetism and arises from dipolar interaction between the two spins in the NV triplet. 
We also consider the Lindbladian operator $\mathcal{L}$ which takes into account the optical pumping rate $\gamma_{las}$ in the $\ket{0}$ state, the longitudinal decay rate $\Gamma_1$ and the spin dephasing rate $\Gamma_2^*$ of the NV$^-$ center.  

Note that we treat the dephasing $\Gamma_2^*$ as a markovian dephasing in the Lindblad formulation. The timescales of the experiments are much slower than the time it takes for the paramagnetic spin bath to change configurations so this does not deeply impact the physics.
Strictly speaking, one should however consider an inhomogeneous gaussian distribution of frequencies due to the slowly fluctuating spin $P_1$ bath. Although we use lorentzian profiles in this description, we chose to keep the $\Gamma_2^*$ notation. 

The Hamiltonian $\hham_0$ and the Lindbladian operators are rotationally invariant around $\bm{e_z}$ which implies that the magnetic susceptibility tensor also verifies this property. We then have:

\begin{align}
\underline{\underline{\chi}}^{NV}=\begin{pmatrix} \chi_{\perp} & -\chi_d & 0\\ \chi_d  & \chi_{\perp} & 0 \\ 0 & 0 & \chi_\parallel \end{pmatrix}.
\end{align}

A small magnetic field perturbation along the NV axis does not affect the eigenvectors of the Hamiltonian neither does it affect the population in these states so that there is no induced magnetization. Thus, $\chi_\parallel=0$. We now evaluate $\chi_d$ and $\chi_{\perp}$. To do so, we consider a small magnetic perturbation along the $\bm x$ direction. We have $\hham=\hham_0+\hat{V}$ with:
\begin{align}
\hham_0 & =\hbar D\hat{S}_z^2+\hbar \gamma_e  B_0 \hat{S}_z \quad {\rm and} \quad
\hat{V}  =\hbar \gamma_e  \delta B_x \hat{S}_x.
\end{align}
The stationary von Neumann Master equation reads:
\begin{align}
0=-\frac{i}{\hbar} [\hham,\hat{\rho}]+\mathcal{L}(\hat{\rho}) \quad {\rm with} \quad \hat{\rho}=\begin{pmatrix} \rho_{11} & \rho_{10} & \rho_{1-1}\\ \rho_{10}^* & \rho_{00} & \rho_{0-1}  
\\ \rho_{1-1}^* & \rho_{0-1}^* & \rho_{-1-1}\end{pmatrix} 
\end{align}
We obtain the set of equations:
\begin{align}
0 & = - i \frac{\gamma_e  \delta B_x}{\sqrt{2}} (\rho_{10}^*-\rho_{10})-\Gamma_1 (\rho_{11}-\rho_{00})-\gamma_{las}\rho_{11}\\
0 & = - i \frac{\gamma_e \delta B_x}{\sqrt{2}} (\rho_{0-1}-\rho_{0-1}^*) -\Gamma_1 (\rho_{-1-1}-\rho_{00})-\gamma_{las}\rho_{-1-1}\\
0 & =-i\left( \frac{\gamma_e  \delta B_x}{\sqrt{2}}  (\rho_{-1-1}-\rho_{00})-\Delta_{-1}\rho_{0-1} + \frac{\gamma_e  \delta B_x}{\sqrt{2}} \rho_{1-1} \right) - \Gamma_2^* \rho_{0-1}\\
0 & =-i\left(\frac{\gamma_e  \delta B_x}{\sqrt{2}}  (\rho_{00}-\rho_{11})+\Delta_{1}\rho_{10} - \frac{\gamma_e  \delta B_x}{\sqrt{2}} \rho_{1-1} \right) - \Gamma_2^* \rho_{10}\\
0 & =-i \left( 2\gamma_e  B_0 \rho_{1-1} + \frac{\gamma_e  \delta B_x}{\sqrt{2}} (\rho_{0-1}-\rho_{10})\right)- \Gamma_2^* \rho_{1-1}\\
1 & =\rho_{11}+\rho_{00}+\rho_{-1-1}
\end{align}

By solving these equations to first order in $\delta H_x$, we get:

\begin{align}
\rho_{11} &= \frac{\Gamma_1}{3\Gamma_1+\gamma_{las}}+o(\delta B_x)\\
\rho_{-1-1} &= \frac{\Gamma_1}{3\Gamma_1+\gamma_{las}}+o(\delta B_x)\\
\rho_{00} &= \frac{\gamma_{las}+\Gamma_1}{3\Gamma_1+\gamma_{las}}+o(\delta B_x)\\
\rho_{0-1} &=\frac{\gamma_{las}}{3\Gamma_1+\gamma_{las}}\left(-\frac{\Delta_{-1}}{\Delta_{-1}^2+(\Gamma_2^*)^2}+i\frac{\Gamma_2^*}{\Delta_{-1}^2+(\Gamma_2^*)^2}\right)\frac{\gamma_e }{\sqrt{2}} \delta B_x +o(\delta B_x)\\
\rho_{10} &=\frac{\gamma_{las}}{3\Gamma_1+\gamma_{las}}\left(-\frac{\Delta_{1}}{\Delta_{1}^2+(\Gamma_2^*)^2}-i\frac{\Gamma_2^*}{\Delta_{1}^2+(\Gamma_2^*)^2}\right)\frac{\gamma_e}{\sqrt{2}} \delta B_x +o(\delta B_x)\\
\rho_{1-1} &=o(\delta B_x)
\end{align}
We can calculate the magnetic moment of one NV$^-$ center via the formula:
\begin{align}
    m_x & =-\hbar \gamma_e \langle \hat{S}_x \rangle =-\hbar \gamma_e \frac{\rho_{0-1}+\rho_{0-1}^*+\rho_{10}+\rho_{10}^*}{\sqrt{2}}\\
    m_y & =-\hbar \gamma_e \langle \hat{S}_y \rangle =-\hbar \gamma_e i \frac{\rho_{0-1}-\rho_{0-1}^*+\rho_{10}-\rho_{10}^*}{\sqrt{2}}\\
m_z & =-\hbar \gamma_e \langle \hat{S}_z \rangle =-\hbar \gamma_e(\rho_{11}-\rho_{-1-1})
\end{align}
To first order in $\delta H_x$, we get:
\begin{align}
   \delta m_x & =\frac{\gamma_{las}}{3\Gamma_1+\gamma_{las}}\left(\frac{\Delta_{-1}}{\Delta_{-1}^2+(\Gamma_2^*)^2}+\frac{\Delta_{1}}{\Delta_{1}^2+(\Gamma_2^*)^2}\right)\hbar \gamma_e^2  \delta B_x \\
   \delta m_y & =\frac{\gamma_{las}}{3\Gamma_1+\gamma_{las}}\left(\frac{\Gamma_2^*}{\Delta_{-1}^2+(\Gamma_2^*)^2}-\frac{\Gamma_2^*}{\Delta_{1}^2+(\Gamma_2^*)^2}\right)\hbar \gamma_e^2  \delta B_x \\
   \delta m_z & =0
\end{align}
We can finally deduce the expressions for $\chi_\perp$ and $\chi_d$. They read:
\begin{align}
   \chi_\perp & = d \hbar \gamma_e^2 \mu_0 \frac{\gamma_{las}}{3\Gamma_1+\gamma_{las}}\left(\frac{\Delta_{-1}}{\Delta_{-1}^2+(\Gamma_2^*)^2}+\frac{\Delta_{1}}{\Delta_{1}^2+(\Gamma_2^*)^2}\right) \\
   \chi_d & = d \hbar \gamma_e^2 \mu_0\frac{\gamma_{las}}{3\Gamma_1+\gamma_{las}}\left(\frac{\Gamma_2^*}{\Delta_{-1}^2+(\Gamma_2^*)^2}-\frac{\Gamma_2^*}{\Delta_{1}^2+(\Gamma_2^*)^2}\right).
\end{align}

Interestingly, the density matrix approach including off-diagonal dephasing $\Gamma_2^*$ predicts an off-diagonal component in the susceptibility. Note that the second order perturbation theory used in eq. (5) if the main text is only valid away from the GSLAC where this term is negligible. 

\section{ NV and Paul trap parameters} 

The experiment is similar to the one used in \cite{DelordNat_SI}.
The diamond sample was bought from the company Adamas and contains $\approx$ 5 ppm of nitrogen-vacancy centers. The green laser is focused by an objective with a numerical aperture of 0.5. The photo-luminescence (PL) is collected by the same objective, separated form the excitation light using a dichroic mirror and a 532nm notch filter, and detected using a multimode-fiber single-photon avalanche photo-detector (SPCM-AQRH-15 from Perkin Elmer). The objective is slightly defocused in order to illuminate the whole diamond. 
The Paul trap is a ring Paul-Straubel trap with a diameter of approximately 200 $\mu$m. It acts both as a trap through the high voltage (HV) and as a microwave (MW) antenna.
The Paul trap provides angular confinement with librational frequencies ranging from 100 Hz to 1 KHz, depending on the particle shape, size, total surface charges.

Due to an intersystem crossing in the excited state of the NV$^-$ center, the electronic spin can be polarized in the $\ket{m_s=0}$ state \cite{Doherty_SI}.
Under typical laser powers, we can polarize the NV spins to the $\ket{m_s=0}$ state with an efficiency above 50\%. 
Due to dipolar interactions amongst the spins, the longitudinal spin-relaxation rate $\Gamma_1$ was measured to be  $\approx$ 2 kHz. The dephasing rate $\Gamma_2^*=5$ MHz, is mostly due to the coupling between the NV centers and the bath of nitrogen spins (so called $P_1$ centers). 
 The laser polarization rate ranges from 10 to 1000 kHz depending in the position of the lens with respect to the diamond and depending on the density of NV centers at the focal point of the laser \cite{DelordNat_SI} due to the aforementioned dipolar processes.

\section{Mechanically-Detected-Magnetic-Resonance}

\begin{figure}[ht]
\centering
\includegraphics[width=\linewidth]{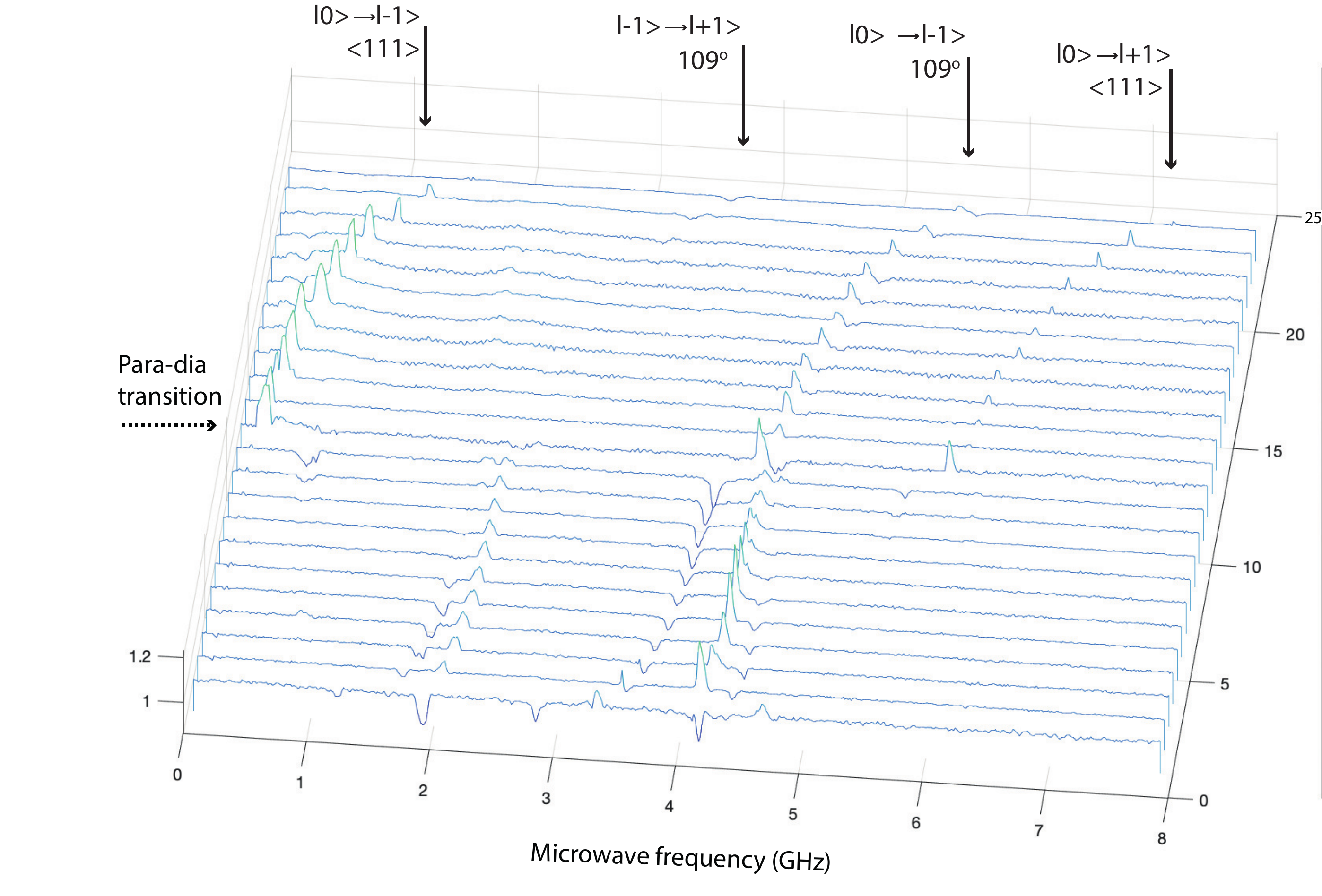}
\caption{Mechanically detected magnetic resonance curves for different magnetic field values. From the position 0 to 25 the magnetic field spans the range 0.03 to 0.16 mT.  In the diamagnetic regime, one sees that the MDMR lines are equidistant because the diamond no longer rotates in this range of magnetic field values. The magnetic resonances are indicated by black arrows.  The states that are coupled are written on the left of the arrows.  $<111>$ indicates the resonance line corresponding to the NV axis that is best aligned with respect to the B field. 109$^\circ$ indicates the microwave transitions for the three NV centers that form an angle of $\approx 109^\circ$ with respect to the best aligned NV centers. 
}
\label{fig2_SI}
\end{figure}

In Mechanically-Detected-Magnetic-Resonance (MDMR) measurements, a microwave resonantly excites the electronic spins of the NV centers, thereby slightly rotating the diamond. Starting from the laser polarized state $\vert m_s=0\rangle$, a microwave creates a net magnetization whose amplitude does not depends on the magnetic field. This is in stark contrast with the induced para/dia-magnetim that we employ to perform large rotations of the diamond crystal.

The diamond motion is detected by collecting the back reflected light from the diamond interface, while scanning the microwave signal. To detect the diamond motion, we focus a small area of the reflected signal onto an optical fibre and detect the photons transmitted through the fibre with a single-photon avalanche photodiode. Compared to the experiment performed in \cite{DelordNat_SI}, we use of an extra laser at 747 nm in order to read-out the angular motion while not impacting the spin-state. This means that the Paul trap angular motion can be estimated without the spin torque.  Using this detection method, motional frequencies can be observed by sending the detected signal to a spectrum analyzer. Under vacuum (1 mbar), the power spectral density (PSD) exhibits narrow peaks at the motional frequencies which are driven by the collisions with the background gaz (Brownian motion).

Thanks to the very high spin-torques and the high angular sensitivity of the speckle detection \cite{DelordNat_SI}, the microwave current enters the Paul trap wire without pre-amplification in these measurements. The microwave power that is employed is around -10 to 0 dBm, as measured before the trap. 
Importantly, only the magnetization component along $\bm{e_z}$ couples to the angular degree of freedom during microwave excitation. The other transverse components of the spin operator oscillate at a frequency greater than 100 MHz, which is 5 orders of magnitude larger than the librational frequency. It thus has no direct impact on the libration.

%In this study, MDMR is only used to use the a slight magnetization along $z$ to track the resonances of the NV centers under the action of the induced para-or dia-magnetism. 
\section{NV Angle and magnetic field read-out} 

The magnetic field amplitude as well as the angle $\theta$ of the NV centers with respect to the B field were estimated using NV magnetometry.
%We study the modification of the angle of the NV centers axis that is closer to the B field axis as a function of the magnetic field amplitude.
Recording the frequency of two spectral lines $(\nu_{0,-1},\nu_{0,+1})$ from the same NV class is enough to estimate both $\theta$ for this NV class as well as the magnitude of $\bm B$. 
This is done by diagonalizing the hamiltonian of the NV electronic spin. The transitions energy between the eigenstates is evaluated numerically for different angles and magnetic fields and
one pair of transition frequency $(\nu_{0,-1},\nu_{0,+1})$ can then be associated to one pair $(\theta,B)$ \cite{HuilleryCPT_SI}.
The width of the MDMR peaks varies from 5 to 20 MHz under weak microwave driving, which translates to errors in the determination of the angle of about 1 degree. 

In regions 1 and 2 of Fig. 2 in the main text, the evolution of the angle $\theta$ that is close to the B field, with respect to the B field amplitude results from a competitions between the torque applied by the Paul trap on the diamond and the other three classes of NV centers.
The characteristic librational frequencies of the Paul trap are 500 Hz - 1kHz. In the region 1, the paramagnetic susceptibility is $\approx 10^{-5}$ implies anti-confinement frequencies that are in the Hz range at these B field values, which is too small to make the spin-torque counteract the Paul trap confinement. In region 2 however the paramagnetic susceptibility is large enough to displace the angle. The dependence of $\theta$ with the magnetic field amplitude in region (2) is particle dependent, as it depends on the angles between the NV orientations and the diamonds axes that are confined by the Paul trap. The sharp drop of the angle from (2) to (3) is however observed at around the same magnetic field value on almost all the particles that we trap. 
%The latter transition is expected to take place at B=102.4 mT, which is very close to the observed critical B-field ($B_c \approx$ 105 mT). 

\section{Reversed bistability} 

 \begin{figure}[ht]
\centering
\includegraphics[width=\linewidth]{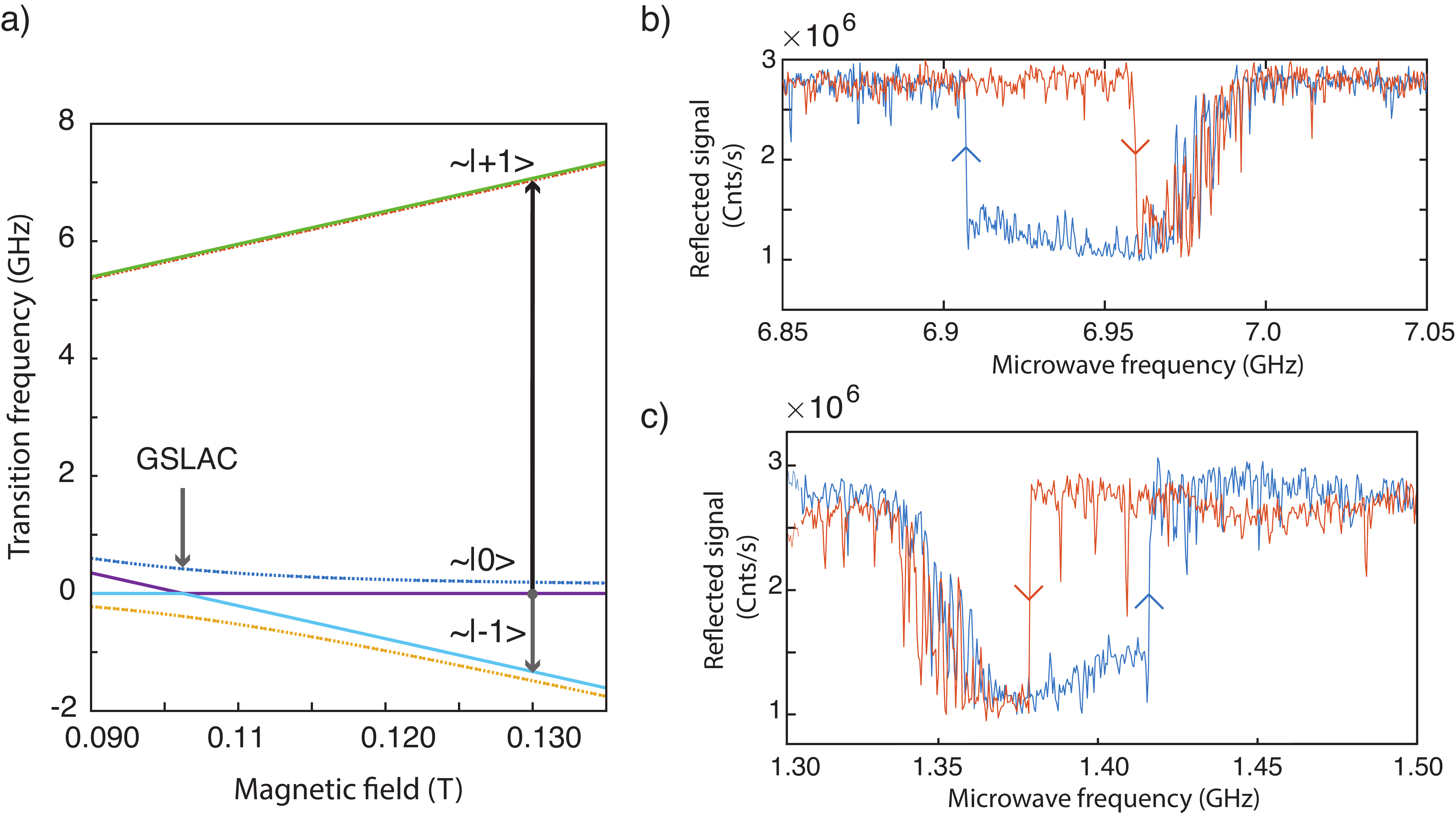}
\caption{a) Evolution of the energies of the three eigenstates in the range 0.09 to 0.135 T for an angle of 0.2 radians (dashed lines) and a perfect alignment (plain lines) between the NV center and the magnetic field. 
 b) Single shot (not averaged) MDMR in the non-linear regimes for the $\vert 0 \rangle$ to $\vert +1 \rangle$ transition when the microwave is scanned from the blue to the red side (blue curve) and from the red to the red side (red curve). When the $\vert +1 \rangle$ state is excited, the angle increases and the transition frequency decreases so the sharp edge is on the red side.   
 c) Single shot MDMR in the non-linear regimes for the $\vert 0 \rangle$ to $\vert -1 \rangle$ transition. Similar color code as in b). Here, when the $\vert -1 \rangle$ state is excited, the angle increases and the transition frequency increases so the sharp edge is on the blue side. }
\label{fig1_SI}
\end{figure}

Before the GSLAC, in the $\vert -1 \rangle$ resp. $\vert +1\rangle$ states, the NV electronic spin states seek low transition frequencies
because the spin torque tends to align/anti-align the NV centers to the B field. 
A consequence of this is that, in the non-linear regime, the sharp edge of the MDMR always appears on the red side \cite{DelordNat_SI}.
In contrast, after the GSLAC, the spins are already aligned to the magnetic field due to the spin-induced diamagnetism, implying a modified non-linear MDMR response. 
The calculation of the energies of the new eigenstates close to the GSLAC is presented in Fig. \ref{fig1_SI}-a) (SM). 
The transverse component of the magnetic field generates an avoided crossing between the bare eigenstates of the $\hat S_{z'}$ operator. 
At sufficiently large laser power and in the dispersive limit (where the states are close to the eigenstates of $S_z$), the intersystem crossing in the optically excited state however polarizes the NV centers in the ground state before the GSLAC and in the second excited state after the GSLAC. When the NV centers are in the $\approx \vert 0 \rangle$ state, after the GSLAC, the spin-mechanical energy is confining. Indeed, any angular motion shifts the energy of the $\vert 0 \rangle$ state upwards.

The situation reverses in the presence of a microwave drive on the $\vert 0 \rangle$ to $\vert -1 \rangle$ transition. 
After the GSLAC, the magnetic energy decreases with angle when the electronic spin is in the $\vert -1 \rangle$ state so the spin misaligns the NV axis from to the B field axis. 
Driving the $\vert 0\rangle$ to $\vert -1 \rangle$ transition in an MDMR means that the particle will then seek large angles.
Due to the avoided crossing, larger angles correspond to a larger transition frequency. In the non-linear regime, the MDMR will thus have their sharp edges on the high frequency side of the MDMR peak. This is what is shown in the inset of Fig. 1-c) in the main text where MDMR was performed with 10 averages. 
Similar measurements are shown in the data of Fig. \ref{fig1_SI}-b) (SM) but with a single shot measurement, enabling sharper jumps to be observed. 

 \begin{figure}[htbp]
\centering
\includegraphics[width=4in]{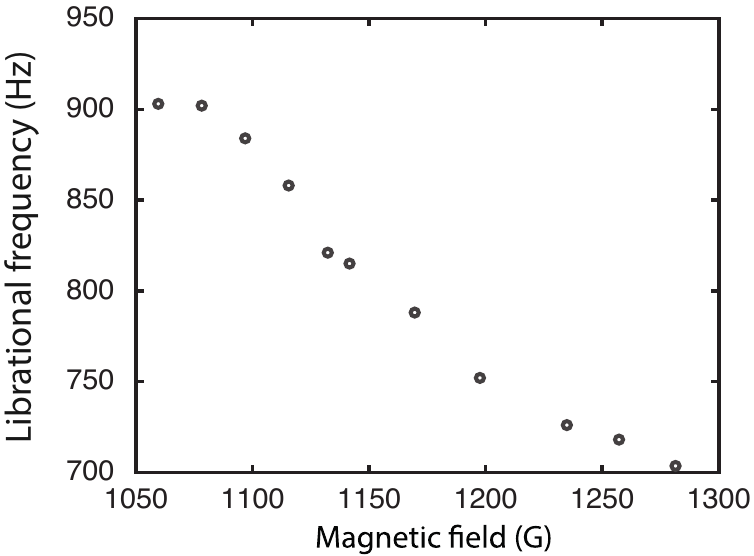}
\caption{Angular confinement characterized by the librational frequency, as a function of the magnetic field.  }
\label{fig3_SI}
\end{figure}

\section{Angle and librational frequency dependency with laser power and magnetic field}

The experimental results shown in Fig. 2 in the main text was realized by recording several MDMR. They are shown in Fig. \ref{fig2_SI} (SM).
The magnetic field amplitude was tuned using a permanent magnet held to a motorized translation stage that displaces the magnet position under vacuum. The angle between the diamond and the magnetic field can be easily determined with MDMR due to the presence of different NV orientations. The ESR of the almost perfectly aligned NV centers exhibits two lines corresponding to the $\ket{0}$ to $\ket{-1}$ and $\ket{+1}$ states. Two magnetic resonance of the three other NV center classes which are misaligned by an angle of $109^\circ$ with the magnetic field are also visible in Fig. \ref{fig2_SI} (SM). Despite the poor population difference between the eigenstates for these NV center classes, those transitions are nevertheless detectable with MDMR due to the large exerted torque under these large applied B fields. For the experiments shown in Fig. 2 in the main text, the translation stage was positioned on a rotating platform that was actuated by hand.
%Fig. 3 of the main text shows the changes in the librational frequency and angles as a function of the magnetic field and laser power. 
%Note that close to the GSLAC, the magnetization along $z$ and the higher order terms in the angle cannot be neglected. 
%These contribute a regularization of the susceptibility which tends to plateau close to 105 mT. 
The parameters with the largest uncertainties for these experiments and those presented in Fig. 3 in the main text are the number of spins $N\approx 10^9$, the moment of inertia $I \approx 10^{-22}$ kg.m$^2$, the Paul trap angular potential depth $U\approx 10^{-15}$J, and laser polarization rate which ranges from 10 to 500 kHz depending on the longitudinal $T_1$ time. The latter depends on the concentration of NV centers, on the so-called ``fluctuators" in the diamond as well as on the angle between the B field and the NV axis. Only the angle of the NV axis and the B field are known with high precision thanks to NV magnetometry.
Using the mean values of the above parameters and a numerical calculation gave agreements by about one order of magnitude with the experimental results that are presented in Fig. 2 and 3 in the main text.
Full numerical fitting using five independent parameters using a three-level model for the spin takes more than one hour on a fast computer. 
Qualitative agreement could be realized using a mixture between a full 7-level models to derive the density matrix and then an analytical calculation. 
Analytical solutions can be found only when the angles between the NV and B field axis are less than 1 degree.
It would however require the Paul trap restoring torque to be orders of magnitude smaller than the spin restoring torque. 

One can also estimate the librational frequency in our experimental conditions (where $\Delta\gg \Gamma_2^*$ and $\Delta \ll D+\gamma_e B$), including the prefactor coming from non-ideal laser polarization. We have 
\be
\omega_\theta= \sqrt{\frac{\hbar N }{I \Delta }   \frac{\gamma_{\rm las}}{3\Gamma_1+\gamma_{\rm las}}}     \gamma_e B.
\ee
$\omega_\theta$ is thus proportional to the magnetic field, as expected for induced magnetic processes.
Taking $B\approx$~0.2~T and $N=10^9$ fully polarized spins in a 15$\mu$m diamond with a moment of inertia $I\approx 10^{-22}$ kg.m$^2$, we obtain
$
\frac{\omega_\theta}{2\pi}\approx 2~\rm kHz,
$
which is greater than the typical Paul trap frequencies, which range from 100 to 1 kHz.

As already mentioned, the above calculations describe the para/dia transition and the resulting confinement in the diamagnetic regime, but cannot be used {\it in extenso} to fit our experiments.
In particular, because of the Paul trap restoring force, the angle between the NV centers and magnetic field is indeed often above 1 degree. A comparison between the analytical calculation and a full numerical treatment shows that, in this regime, the analytical theory must be modified to go beyond the linear approximation and include all three levels.
Unfortunately, the Paul trap torque also prohibits studying the predicted non-diagonal component of the susceptibility. 
Lowering the Paul trap confinement means reducing the center of mass frequency so that the particle leaves the RF null region of the trap. 

The inset of Fig. \ref{fig3_SI} (SM) shows the PSD zoomed to one librational mode as a function of the magnetic field.
The central frequency of this mode decreases as a function of the magnetic field in qualitative agreement with the above analysis.

 \begin{figure}[htbp]
\centering
\includegraphics[width=4in]{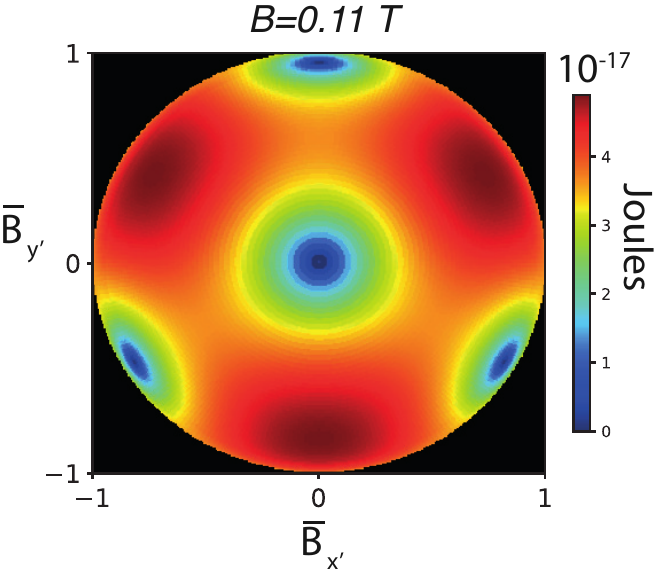}
\caption{Magnetic energy as a function of the magnetic field directions after the GSLAC and with $10^9$ NV centers. $\overline{B}_{x',y'}= {B}_{x',y'}/\vert \bm B\vert $, where $x'$ and $y'$ are the directions perpendicular to the central NV axis. }
\label{fig4_SI}
\end{figure}

\section{Magnetic potential energy} 

The potential energy coming from the electronic spins of the NV centers in the magnetic field are derived by computing the mean of the torque operators, defined as $\hat \tau_\theta=-\frac{\partial \hham}{\partial \theta}$ for the angle $\theta$ and then integrating over the angles $\theta$ and $\phi$ to obtain the potential energy.
This potential landscape is plotted in Fig. \ref{fig4_SI} (SM) for a magnetic field of $0.11$T.

The Fig. \ref{fig5} (SM) is plotted for the angle $\phi$ equal to 0 degree. For this value of $\phi$, the magnetic energy is not a symmetric function of $\theta$. This is due to the three other NV classes which break the rotational invariance around a single NV center class.

We see that the magnetic energy minimum tends to get closer to the angle $\theta=0^\circ$ by increasing the value of the magnetic field. This explains why the diamond rotates continuously towards region (3) in the Fig. 2 of the main text. After the GSLAC, the magnetic energy minimum is reached when the NV is perfectly aligned with the magnetic field.  Before this the diamond is weakly bound to a region where the angle $\theta$ can be large. 

 \begin{figure}[htbp]
\centering
\includegraphics[width=4in]{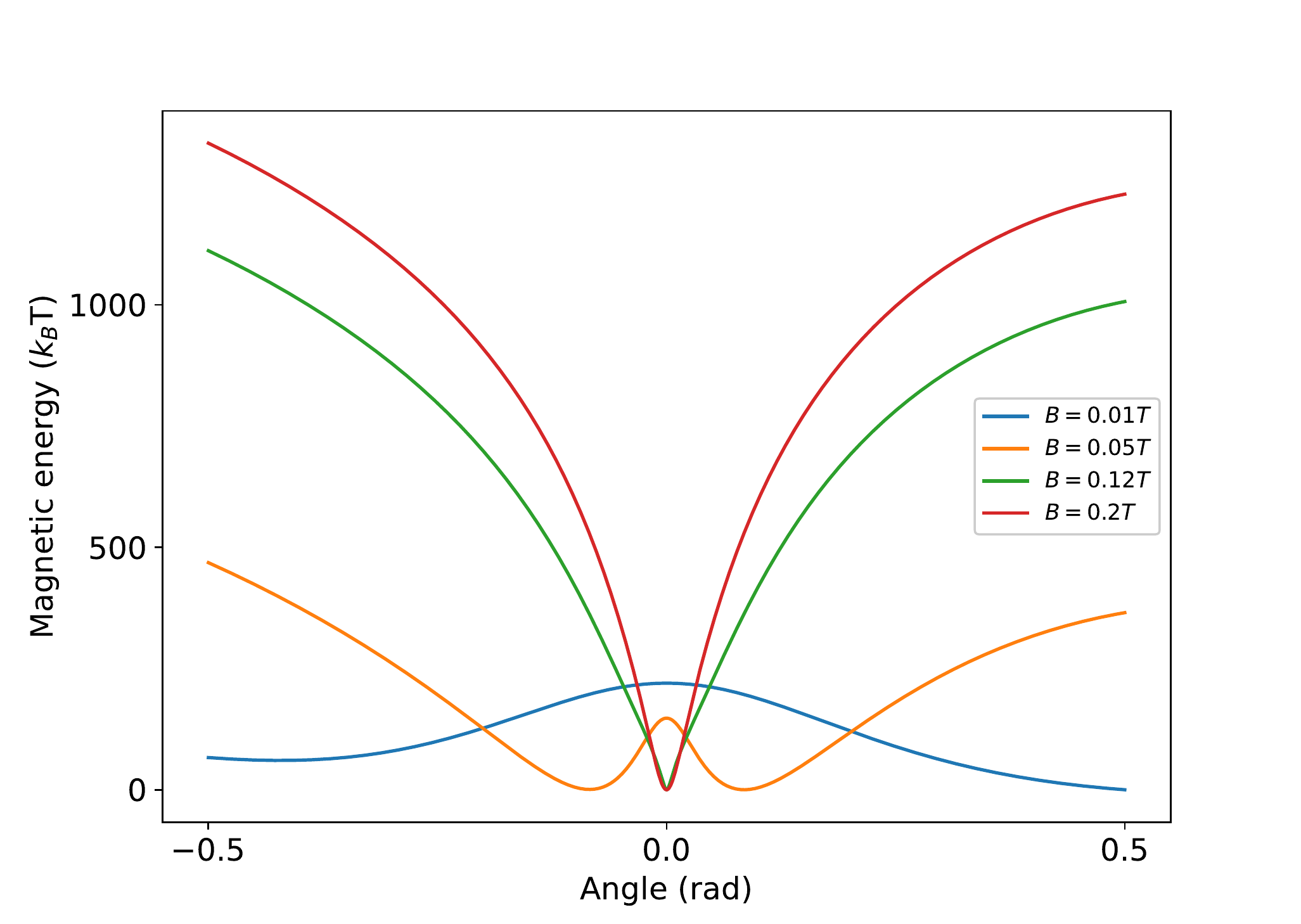}
\caption{Magnetic energy as a function of the angle $\theta$ for different values of the external magnetic field with a diamond containing $10^9$ NV centers.}
\label{fig5}
\end{figure}

\end{widetext}

\end{document}